\documentclass[AMA,STIX1COL]{WileyNJD-v2}

\usepackage{siunitx}
\usepackage{tabulary}
\usepackage{tabularx}
\usepackage{longtable}
\usepackage{setspace}
\usepackage{enumitem}
\usepackage{multirow}
\usepackage{graphicx}
\usepackage{colortbl}

\usepackage{array}
\usepackage{lipsum}
\usepackage{mwe}
\usepackage[space]{grffile}  
\usepackage{arydshln}   
\usepackage{listings}
\usepackage{verbatim}
\usepackage{bbm}
\newcolumntype{L}[1]{>{\raggedright\arraybackslash}p{#1}}
\newcommand{\noteC}[1]{{\footnote{\color{red} \bf{from Cory:} #1} }}

\usepackage{amsmath,amsfonts,amssymb}

\articletype{Original Article}

\received{13 June 2018}
\revised{15 February 2019}

\begin{document}

\title{Uncertainty in the Design Stage of Two-Stage Bayesian Propensity Score Analysis}


\author[1]{Shirley X. Liao PhD}
\author[2]{Corwin M. Zigler PhD}


\address[1]{\orgdiv{Department of Biostatistics}, \orgname{Harvard School of Public Health}, \orgaddress{\state{MA}, \country{USA}}}

\address[2]{\orgdiv{Department of Statistics and Data Sciences}, \orgname{The University of Texas at Austin}, \orgaddress{\state{TX}, \country{USA}}}


\corres{Shirley X. Liao PhD, Harvard School of Public Health, 677 Huntington Ave, Boston, USA
\email{shirleyxliao@g.harvard.edu}}



\authormark{LIAO AND ZIGLER}


\abstract[Summary]{The two-stage process of propensity score analysis (PSA) includes a design stage where propensity scores are estimated and implemented to approximate a randomized experiment and an analysis stage where treatment effects are estimated conditional upon the design. This paper considers how uncertainty associated with the design stage impacts estimation of causal effects in the analysis stage. Such design uncertainty can derive from the fact that the propensity score itself is an estimated quantity, but also from other features of the design stage tied to choice of propensity score implementation. This paper offers a procedure for obtaining the posterior distribution of causal effects after marginalizing over a distribution of design-stage outputs, lending a degree of formality to Bayesian methods for PSA (BPSA) that have gained attention in recent literature. Formulation of a probability distribution for the design-stage output depends on how the propensity score is implemented in the design stage, and propagation of uncertainty into causal estimates depends on how the treatment effect is estimated in the analysis stage. We explore these differences within a sample of commonly-used propensity score implementations (quantile stratification, nearest-neighbor matching, caliper matching, inverse probability of treatment weighting, and doubly robust estimation) and investigate in a simulation study the impact of statistician choice in PS model and implementation on the degree of between- and within-design variability in the estimated treatment effect. The methods are then deployed in an investigation of the association between levels of fine particulate air pollution and elevated exposure to emissions from coal-fired power plants.

}


\maketitle

\footnotetext{\textbf{Abbreviations:} PSA, propensity score analysis; BPSA, Bayesian propensity score analysis; NN, nearest neighbor; DEU, design estimation uncertainty; DDU, design decision uncertainty; AEU, analysis estimation uncertainty}

\section{Introduction} \label{intro}

Propensity score (PS) analysis refers to a wide range of strategies for estimating causal treatment effects with observational data. Rubin and others \cite{rubin, hernan_invited_2005} motivate this process by conceptualizing an observational study as having arisen from a randomized clinical trial where the rules of assignment have been lost and must be estimated. This perspective draws focus to the ``design'' stage of a propensity score analysis (PSA), in which PS are estimated then implemented to create a sub- or pseudo-population of the data representative of a hypothesized randomized trial \cite{rubin,harder_propensity_2010}.  Following the design stage, observed outcome information is used to estimate a treatment effect in the ``analysis'' stage. 

 A well-conducted design stage is ``absolutely essential for drawing objective inferences for causal effects,'' and is typically conducted without access to outcome data \cite{rubin}.  Many analytic decisions are required to create a successful design which implements the estimated propensity score (e.g., via matching, weighting, or subclassification) to achieve treated and untreated groups that are ``balanced'' with respect to observed covariates \cite{fanli,alvarez,cefalu,stuart,austin}. In traditional PSAs, once the researcher has satisfactorily approximated the design of the hypothesized randomized study, the sub- or pseudo- population of observations created in the design stage is treated as fixed or known and estimation of causal effects in the analysis stage is conducted conditional on the design. In light of the importance of the design stage and the multi-faceted decisions made towards estimating and implementing the propensity score, it stands to reason that uncertainty associated with the design stage of PSA should be propagated into the estimation of causal effects in the analysis stage in order to more fully acknowledge all potential sources of PSA uncertainty.

Consideration of design uncertainty is often framed solely in terms of whether and how to incorporate uncertainty from the estimation of the PS into variance estimation of causal effects \cite{Ho, wooldridge_inverse_2002, brumback_note_2009, liangli, caliendo, hirano, graham}.  However, more sources of uncertainty may exist in the design stage.  Dehejia and Wahba \cite{dehejia} illustrated that when implementing the PS with nearest neighbor matching without replacement, the ordering of observations in the matching procedure resulted in different estimates of treatment effect, even when using the same set of estimated PS \cite{dehejia}. Abadie and Imbens derive a variance estimator for causal effects that accounts for both PS estimation uncertainty and uncertainty in the construction of matches \cite{abadie}.  Further expansion of the idea of design uncertainty has appeared in Zigler and Dominici, who discuss uncertainty in choice of covariates included in the PS model \cite{zandd}, Spertus and Normand \cite{Spertus} who construe probabilistic design weights, and Zigler and Cefalu, who regard the subset of observations pruned (or truncated) from the design to be unknown \cite{cefalu}.

Rather than conduct estimation of causal effects conditional on a single design stage, we offer an analysis that first generates a probability distribution for the space of all possible designs, the structure of which is determined by a) the PS estimation model and b) the specific type of implementation used in the design. To propagate design uncertainty into effect estimation, inference in the analysis stage is then marginalized over the distribution of possible designs while maintaining separation between the outcome data and the design stage.  Developing a distribution for the space of designs clarifies the distinction between quantities we define as 1) ``design estimation uncertainty'' (DEU) corresponding to familiar notions of estimation uncertainty in the PS, and 2) ``design decision uncertainty,'' (DDU) corresponding additional stochasticity inherent to some PS implementations which lead to various design outputs for a fixed value of the PS.  Uncertainty arising from treatment effect estimation in the analysis stage is defined as analysis estimation uncertainty (AEU). These elements of uncertainty are considered in a setting where the following PSA decisions are made \textit{a priori}: the causal estimand of interest, the form of the PS model, the type of PS implementation (e.g., matching, weighting, etc.), and the method of treatment effect estimation in the analysis stage.  

The method described herein is operationalized with the mechanics of Bayesian inference.  While performing Bayesian inference separately in either the ``design'' or ``analysis'' stage is straightforward, combining them in a single Bayesian updating approach requires careful consideration.  Bayesian inference that maintains the inherent two-stage nature of PSA gained attention beginning with McCandless \cite{mccandless_bayesian_2009, mccandless_cutting_2010}, and continued to appear in following literature \cite{an_bayesian_2010, kaplan, zigler_model_2013, zigler_uncertainty_2014, zigler_central_2016}.  These procedures fall within a broader class of methods for modularized Bayesian inference \cite{liu_modularization_2009, jacob_better_2017}, where a ``model made of modules"  may restrict the propagation of information from one module to another for reasons such as preventing the misspecification of one module from contaminating another, although there are other motivating considerations \citep{jacob_better_2017}.  In the PS context, the design and analysis stages can be cast as different modules, with care taken so that the analysis does not contaminate the design.  Such restricted flow of Bayesian updating has also been described as a ``cut" or a ``valve" to prevent feedback of information \cite{plummer, lunn_combining_2009}.  One goal of this paper is to bring together many such issues that have appeared sporadically in literature towards improved understanding of the role of Bayesian methods for accounting for different elements of uncertainty in a propensity score analysis.  As will be elaborated, deploying the mechanics of Bayesian inference across distinct ``design'' and ``analysis'' stages (or modules) has close ties to multiple imputation (MI) of missing data.  As in MI, Bayesian ideas motivate considerations for marginalizing over uncertainty in the unknown design, but notions of formal Bayesian validity are elusive in this instance, and thus the procedures contained herein are evaluated on the basis of their ``calibration'' or Frequentist operating characteristics \citep{little_calibrated_2011, murray}.

We examine this framework within the context of several commonly-used PS implementations and corresponding analysis stages: quantile stratification, nearest-neighbor matching with replacement, caliper matching with replacement, inverse probability of treatment weighting (IPTW), and a common doubly-robust estimator, although the formalization applies to a broader range of PSAs.  We consider both settings where the analysis stage consists of a parametric ``outcome model'' (conditional on the PS), and also settings where no outcome-model likelihood exists, as would be the case with many common non- or semi-parametric (e.g., matching or weighted) estimators with known asymptotic properties.  The latter case is particularly complicated for Bayesian inference, in which case we propose an approximation to the asymptotic posterior distribution of a causal effect.  The collection of PS analyses considered here is not designed to be exhaustive and we make no argument for superiority of any given approach.  Rather, these specific implementations and analysis stages are used to illustrate the different mechanistic sources of uncertainty that can arise when performing a PSA. 

After distinguishing between different sources of design uncertainty and offering a formalization of integrating design uncertainty into the analysis stage, we outline a corresponding computational algorithm and illustrate the approach in a simulation study which demonstrates how the quantity of design uncertainty varies under choices a statistician makes in the design stage - from the variables included in the PS model to the form of PS implementation utilized. Arguments borrowed from MI literature will be used to compare between- and within-design variability in treatment effect estimates in order to quantify how design uncertainty impacts posterior inference on treatment effects. We then illustrate this method in an analysis of emissions from coal-fired power plants and ambient particulate air pollution collected over 22,723 zip codes in the Northeast, Southeast and Industrial Midwest regions of the United States.   The paper concludes with a discussion of possible future directions.


\section{Notation, Estimands, and Overview of Marginalizing over Design Uncertainty} \label{basics}

\subsection{Notation and Estimand}\label{basics:notation}
Let $Y_i$, $X_i$ and $T_i$ represent the observed outcome, covariate vector (of length $p$) and treatment indicator for unit $i$, where $i = 1,2,\ldots, n$. $T_i \in [0,1]$ is dichotomous and $Y_i$ is a continuous random variable. Let $\boldsymbol{Y}$ and $\boldsymbol{T}$ be vectorized representations of the data and $\boldsymbol{X}$ a matrix of observed covariates. We state without comment the following assumptions common to causal inference literature and defer readers to other work for details\cite{rubin}: positivity (each subject has a non-zero probability of receiving either treatment) and SUTVA (units do not interfere with each other's outcomes and potential outcomes are well-defined). 

The additional key assumption of strongly ignorable treatment assignment states that potential outcomes which exist under either treatment are independent of observed treatment assignment after adjusting for observed covariates in \textbf{X}, which must include all confounders.  Under ignorability, the average treatment effect (ATE) may be estimated from observed outcomes among units that exhibit the same distribution in background covariates:

\begin{equation}
\Delta_{ATE} = E[E(Y_i|T_i = 1,X_i) - E(Y_i|T_i = 0,X_i)]
\end{equation}

All relevant covariate information may be summarized in a ``coarsened'' manner with the propensity score, defined as the probability to receive the treatment rather than the control conditional on pre-treatment covariates of that individual \cite{rosenbaum2}. Let $e_i$ represent the propensity score for individual $i$.

\begin{align}
e_i = P(T_i = 1|X_i)
\end{align}

Units which are homogeneous with respect to $e_i$ are said to be ``balanced'' with respect to $X_i$, and under the assumption of strong ignorability, units with the same value of propensity score but assigned to different treatments have an expected difference in responses equal to the average treatment effect \cite{rosenbaum2}.

While $\Delta_{ATE}$ is defined above as the marginal average treatment effect in the population, $\Delta$ (without a subscript) will be used from this point to generically represent a causal estimand such as the ATE or the  ATT(C) (Average Treatment Effect on the Treated (Control)). The choice of estimand considered is implied by the PS implementation, which is assumed to be chosen {\it a priori}. Differences between estimands are not the focus of this paper, rather, we will consider estimation of each in the context of a corresponding implementation stage (e.g., matching will be implicitly taken to be estimating the ATT, whereas IPW estimates the ATE).

\subsection{Overview of Marginalizing over Design Uncertainty} \label{basics:overview}

We begin with a heuristic description of the proposed framework for Bayesian PSA, deferring details to subsequent sections.  Without the use of the propensity score, traditional Bayesian inference for $\Delta$ would follow from specification of a likelihood for $(\boldsymbol{T,X,Y})$ conditional on unknown parameters, $\theta$, a prior distribution for $\theta$, and some function relating the data and $\theta$ to the quantity $\Delta$.  However, this standard procedure for Bayesian inference to jointly estimate the propensity score and $\Delta$ does not provide reliable causal estimates, as the PS is not intended to be part of the data-generating likelihood for $\boldsymbol{Y}$ \cite{zigler_central_2016, robins_discussion_2015, robins_toward_1997}, thus motivating careful delineation of unknown quantities in the design and analysis stages. 

Let $\nu$ be a parameter summarizing the output of a PS implementation, that is, representing one propensity score ``design'' implied by implementing a set of estimated propensity scores.  For example, $\nu$ may be defined as a partition of observations into strata based on their estimated propensity scores (detailed definition of $\nu$ is developed in the context of several PS implementations in Section \ref{ps.design:imp}). Under the assumptions outlined in Section \ref{basics:notation} and a design stage that is successful in the sense that it appropriately balances observed covariates between treated and untreated units, inference for $\Delta$ may be carried out with standard methodology (Bayesian or otherwise) conditional on a single value of $\nu$, possibly with variance estimators adapted to account for uncertainty in $\nu$ \cite{rubin_matching_1996, abadie, wooldridge_inverse_2002}. Instead of conditioning on a single realization of $\nu$, the goal of this paper is to marginalize posterior inference for $\Delta$ over the space of reasonable values of $\nu$, each with an associated probability. This is represented heuristically as:

\begin{align}\label{eq:overviewexpression}
f(\Delta|\boldsymbol{T},\boldsymbol{X},\boldsymbol{Y}) = \int_{\nu} f(\Delta|\boldsymbol{T},\boldsymbol{X},\boldsymbol{Y},\nu) f(\nu|\boldsymbol{T},\boldsymbol{X}) d\nu 
\end{align}

The distribution $f(\nu|\boldsymbol{T,X})$ is written to explicitly acknowledge that the design stage is conducted without outcome information ($\boldsymbol{Y}$).  Integrating the posterior distribution of $\Delta$ over the marginal distribution of $\nu$ is a mathematical representation of propagating design uncertainty into the analysis stage by marginalizing over the space of all possible designs. 



Expression (\ref{eq:overviewexpression}) also highlights an important analog to the analysis of missing data:  Construing $\nu$ as a missing quantity closely parallels the Bayesian derivation of a multiple imputation procedure, where ($\boldsymbol{T,X,Y}$) represent the observed data and the expression $f(\nu|\boldsymbol{T,X})$ represents a model for the missing data, intentionally specified in this case to omit $\boldsymbol{Y}$ to preclude the outcome from contributing to the design.  Integration over $\nu$ represents the multiple imputation of ``missing'' designs, with the analysis stage conducted conditional on each and, as will be detailed in the subsequent, uncertainty in the parameters governing the construction of $\nu$ (in addition to the value of $\nu$ itself) being propagated into estimates of $\Delta$.  The modularity of BPSA means that, analogous to procedures for multiple imputation, $f(\Delta|\boldsymbol{T,X,Y})$ may not be regarded as a standard Bayesian posterior \cite{murray, little_calibrated_2011}.  Detailed specification of the modules will follow in Sections \ref{ps.design} and \ref{analysis}, with the corresponding computational procedure described in Section \ref{analysis:computation}.





\section{A Distribution of Designs Governing Design Uncertainty} \label{ps.design}

\subsection{Bayesian Propensity Score Estimation and Design Estimation Uncertainty} \label{ps.design:deu}

The parameter $\nu$ represents the result of propensity score estimation and implementation for which ignorable treatment assignment is assumed, thus its probability distribution $f(\nu|\boldsymbol{X},\boldsymbol{T})$ must be anchored to the PS estimation model. We expand the posterior distribution of $\nu$ to reflect its dependence on parameters from the PS model, which represents the ``design module'' of expression (\ref{eq:overviewexpression}):

\begin{align}\label{eq:deuexpression}
f(\nu|\boldsymbol{X},\boldsymbol{T})  = \int f(\nu|\alpha,\boldsymbol{X},\boldsymbol{T}) f(\alpha|\boldsymbol{X},\boldsymbol{T})d\alpha  = \int f(\nu|\alpha, \boldsymbol{X}, \boldsymbol{T})L(\boldsymbol{T}|\boldsymbol{X},\alpha)\pi(\alpha)d\alpha,
\end{align}
where $L(\boldsymbol{T}|\boldsymbol{X},\alpha)$ represents a likelihood function for the treatment assignment mechanism (i.e., a propensity score model) and $\pi(\alpha)$ represents a prior distribution for the unknown parameter $\alpha$.  

For illustration throughout this paper, we consider a simplistic propensity score model based on a generalized linear model $L(\boldsymbol{T}|\boldsymbol{X},\alpha) = \prod_{i=1}^n g(\alpha X_i)^{T_i} (1 - g(\alpha X_i))^{1 - T_i}$, with a logit link function $g(\alpha X_i)$, although the ensuing discussion will be relevant to any parametric propensity score model specification with unknown parameter $\alpha$.  Note that the propensity score for each observation, $e_i$, is a deterministic function of $(\alpha, \boldsymbol{X}, \boldsymbol{T})$.  In the case of a logistic propensity score model, $e_i = g(\alpha X_i) = \frac{e^{\boldsymbol{\alpha} X_i}}{1+e^{\boldsymbol{\alpha} X_i}}$. A traditional Frequentist estimate of the propensity score may be obtained by replacing the unknown parameter $\alpha$ with its maximum likelihood estimate. Instead, integrating with respect to $\alpha$ in expression (\ref{eq:deuexpression}) may be thought of as simulating from the posterior ``predictive'' distribution of propensity scores for each individual \cite{rosenbaum2}. From this point forward, we refer to $\alpha$ and the PS interchangeably with regards to describing a posterior distribution of quantities derived from the treatment assignment model.

Variability in the predictive distribution of propensity scores captures uncertainty in the estimation of the propensity scores, which we refer to as ``design estimation uncertainty'' (DEU).  Propagation of DEU into causal effect estimates has been previously considered in Bayesian and non-Bayesian contexts \cite{mccandless,alvarez,kaplan}.

\subsection{Propensity Score Implementation and Design Decision Uncertainty}
\label{ps.design:imp}

Define $\nu$ as a vector of length $n$, where $\nu = [\nu_1,\nu_2,\ldots, \nu_n]$ and $\nu_i$ encodes the output of the PS implementation for individual $i$. The parameter space for each $\nu_i$ depends upon the choice of PS implementation procedure and the space of the entire vector $\nu$ may receive additional restrictions based on the details of the implementation.  Delineation of the components of $f(\nu|\boldsymbol{T,X})$ as expressed in expression (\ref{eq:deuexpression}) draws a distinction between uncertainty derived from estimation of the propensity score (DEU) and any additional source of uncertainty owing to the stochasticity inherent to the implementation, to which we refer as ``design decision uncertainty'' (DDU).  The possibility of DDU is denoted by the expression $f(\nu|\alpha, \boldsymbol{X}, \boldsymbol{T})$.


A PS implementation may or may not generate DDU. We refer to implementations that do not generate DDU as {\it deterministic implementations}, which produce only one possible value of $\nu$ for each value of $\alpha$. In these implementations, $f(\nu|\boldsymbol{T,X},\alpha)$ is a point-mass function and all design uncertainty derives from DEU.  In contrast, {\it probabilistic implementations} can produce multiple values of $\nu$ for a single $\alpha$, as dictated by the non-degenerate distribution $f(\nu|\boldsymbol{T,X},\alpha)$. In the case of probabilistic implementations, both DDU and DEU contribute to design uncertainty. 

We consider specifying $\nu$ and $f(\nu|\boldsymbol{T,X},\alpha)$ in the context of five propensity score implementations: quintile stratification, nearest neighbor matching with replacement, caliper matching with replacement, and weighting for normalized inverse-probability-weighted (IPW) and doubly-robust (DR) estimators. 

\subsubsection{Stratification} \label{ps.design:imp:strat}

Stratification is a type of sub-classification performed on quantiles of the estimated propensity score distribution. For this implementation, $\nu_i$ may be specified as a categorical variable, with $\nu_i = q$ if individual $i$ is assigned to strata $q$, when the sample is stratified into one of $q=1,2,\ldots,Q$ quantiles. We consider quintile stratification such that $Q=5$, $\nu_i$ takes on a value in $[1,2,3,4,5]$ and $\nu$ is defined on the space of possible allocations of $n$ units into quintiles. Stratification is a deterministic PS implementation method.


\subsubsection{Nearest Neighbor Matching With Replacement, implemented with a caliper} \label{ps.design:imp:nn}
Nearest neighbor (NN) matching with replacement considers treated observations one at a time and matches each to the control observation(s) (ratio of matching decided \textit{a priori}) with the closest propensity score. Controls are allowed to be matched to multiple treated observations, and thus may appear multiple times in the matched set. By implementing NN matching with a caliper, the pool of possible matches is limited by the caliper width, preventing unsuitable matches and instead pruning treated observations with propensity scores too removed from the distribution of control propensity scores.

In this implementation (and all matching with-replacement implementations), $\nu_i$ may be defined as the frequency weight for observation $i$, which is calculated by standardizing the frequency of inclusion of each observation to sum to the number of unique observations within each treatment arm in the matched set. Thus $\nu_i$ is defined on the space of real numbers bounded above by the maximum (across treatments) number of observations within each treatment arm. 

As the case with stratification, NN matching with replacement is a deterministic implementation yielding one matched set for each set of propensity scores and a point-mass function $f(\nu|\boldsymbol{T,X},\alpha)$.

\subsubsection{Caliper Matching With Replacement}\label{ps.design:imp:caliper}

In addition to NN matching, we also consider matching with replacement utilizing a caliper, which we refer to as ``caliper matching.'' While this implementation is not widely used in propensity score literature, we incorporate it specifically as a stochastic counterpart to the NN algorithm of Section \ref{ps.design:imp:nn}. The difference between the two matching algorithms is that while NN chooses matches deterministically (closest based on PS distance) caliper matching chooses matches randomly among the candidates contained within the caliper. 

Within the context of caliper matching with replacement, $\nu$ is defined in the same manner as NN matching with replacement. However, the random choice of control matches within the caliper renders this a probabilistic implementation; even for a fixed $\alpha$, there may still be variability in $\nu$ owing to the random selection of matches.  $f(\nu|\boldsymbol{T,X},\alpha)$ is a probability distribution without an easily obtainable closed-form solution, but draws may be taken via iteratively performing the caliper matching algorithm multiple times conditional on the same $\alpha$. 

\subsubsection{Weighting for IPW and DR estimators}\label{ps.design:imp:weighting}

In contrast to matching and subclassification, weighting implementations utilize the propensity score to create a ``pseudo-population'' of observations in the treatment group, control group, or both in order to balance covariate distributions between the two groups. Inverse probability weighting (IPW) assigns weights based the following transformation of the propensity score:

\begin{align}\label{eq:weights}
w_i = \frac{T_i}{e_i} + \frac{1-T_i}{1-e_i}
\end{align}

For IPW, $\nu_i = w_i$, living on the space of positive real numbers. Since weights are created from a one-to-one transformation of a given a set of PS, the implementation is deterministic and $f(\nu|\boldsymbol{T,X},\alpha)$ is point-mass.  This applies whether the weights are deployed in a standard IPW analysis or in tandem with an outcome model specification towards construction of a doubly-robust estimator. 

\section{Sampling from the marginal posterior distribution of treatment effects} \label{analysis}

This section puts the distribution of $\nu$ formulated in Section \ref{ps.design} within the modularized Bayesian framework presented in Section \ref{basics:overview} with the goal of obtaining posterior inference on the marginalized posterior distribution of treatment effects. Incorporating the relationship between design output $\nu$ and the PS model, we expand expression (\ref{eq:overviewexpression}) as:
\begin{align} \label{eq:post.delta}
f(\Delta|\boldsymbol{T},\boldsymbol{X},\boldsymbol{Y}) = \int_{\nu} f(\Delta|\boldsymbol{T},\boldsymbol{X},\boldsymbol{Y},\nu) f(\nu|\boldsymbol{T},\boldsymbol{X}) d\nu \\ = \int_{\nu} f(\Delta|\boldsymbol{T},\boldsymbol{X},\boldsymbol{Y},\nu)  \int_{\alpha} f(\nu|\alpha,\boldsymbol{X},\boldsymbol{T}) f(\alpha|\boldsymbol{X},\boldsymbol{T}) d\alpha d\nu \\
= \int_{\nu} 
\underbrace{f(\Delta|\boldsymbol{T},\boldsymbol{X},\boldsymbol{Y},\nu)}_\text{AEU} \int_{\alpha} \underbrace{f(\nu|\alpha,\boldsymbol{X},\boldsymbol{T})}_\text{DDU} \underbrace{\pi(\alpha) \prod_{i=1}^n L(T_i|X_i,\alpha)}_\text{DEU} d\alpha d\nu 
\end{align}

The posterior distribution of treatment effects conditional on the design is denoted with $f(\Delta| \boldsymbol{T}, \boldsymbol{X}, \boldsymbol{Y}, \nu)$, representing the ``analysis module'' of BPSA, with variability in this distribution defined as ``Analysis Estimation Uncertainty'' (AEU). The functional form of $f(\Delta| \boldsymbol{T}, \boldsymbol{X}, \boldsymbol{Y}, \nu)$ is specified {\it a priori}, depending in part on the structure of $\nu$ implied by the design stage.  Some PS implementations permit $f(\Delta| \boldsymbol{T}, \boldsymbol{X}, \boldsymbol{Y}, \nu)$ to be specified based on a parametric outcome model, while others dictate specification of $f(\Delta| \boldsymbol{T}, \boldsymbol{X}, \boldsymbol{Y}, \nu)$ using weighted estimators with known asymptotic properties, which are typically not motivated on Bayesian grounds.  Sections \ref{analysis:parametric} and \ref{analysis:asymptotic} outline development in each case.  



\subsection{Formalizing the conditional posterior distribution of $\Delta$} \label{analysis:formal}
\subsubsection{Analysis stages where $\Delta$ is estimated with a parametric model} \label{analysis:parametric}

We first consider settings where the analysis stage consists of estimating a parametric outcome model, conditioned on $\nu$. Following propensity score stratification, for example, the analysis stage may be performed with a parametric outcome model having the likelihood function $L(\boldsymbol{Y}|\boldsymbol{T},\boldsymbol{X},\boldsymbol{Y},\nu,\theta)$, where $\Delta$ is defined as a function of the data and unknown parameters $\theta$. Allowing definition of the conditional posterior distribution of $\Delta$ from expression (\ref{eq:post.delta}) as:

\begin{align}\label{outcomeposterior}
f(\Delta|\boldsymbol{T},\boldsymbol{X},\boldsymbol{Y},\nu) = \int f(\Delta|\boldsymbol{T},\boldsymbol{X},\boldsymbol{Y},\nu,\theta)f(\theta|\boldsymbol{T},\boldsymbol{X},\boldsymbol{Y},\nu)d\theta
\end{align}

where $f(\theta|\boldsymbol{T},\boldsymbol{X},\boldsymbol{Y},\nu) = L(\boldsymbol{Y}|\boldsymbol{T},\boldsymbol{X},\boldsymbol{Y},\nu,\theta)\pi(\theta)$ is the posterior distribution of the outcome model parameters under prior distribution $\pi(\theta)$.  For illustration, we consider $f(\cdot)$ to be a linear regression model with $\theta = (\beta, \sigma^2)$ representing, respectively, a vector of regression coefficients and a conditional variance parameter. In this paper, we only consider a parametric outcome model in the case of stratification, where $\nu$, a factor variable indicating stratum membership, is included as a covariate in the outcome model, along with an interaction with the treatment. In this case, the causal effect $\Delta$ is a function of $\theta$ and observed data.

\subsubsection{Analysis stages where $\Delta$ is estimated with an estimator with known asymptotic properties} \label{analysis:asymptotic}

Some PS implementations, in particular those based on weighting, do not lend themselves to estimation of $\Delta$ with a parametric outcome model. The form of the estimator depends on the definition of the weights, which we illustrate with implementations of caliper matching with replacement, NN matching with replacement, IPW, and a simple DR estimator (see Appendix \ref{table:ps.imp} for weights and weighted estimators of these implementations). The lack of a parametric model for $\boldsymbol{Y}$ and corresponding likelihood expression requires a different approach for specifying the conditional distribution $f(\Delta|\boldsymbol{T},\boldsymbol{X},\boldsymbol{Y},\nu)$ from expression (\ref{eq:post.delta}).  

One option utilized in past literature \cite{kaplan} is specification of $f(\Delta|\boldsymbol{T},\boldsymbol{X},\boldsymbol{Y},\nu)$ with the known asymptotic distribution of the estimator for $\Delta$, conditional on one realization of $\nu$. Let $\hat{Q}(\nu)$ represent an estimator of $\Delta$ and $\hat{V}(\nu)$ the sampling variance, both computed conditional on design $\nu$. Then an asymptotic approximation takes the form:

\begin{align}
f(\Delta|\boldsymbol{T},\boldsymbol{X},\boldsymbol{Y},\nu) \dot\sim N(\hat{Q}(\nu),\hat{V}(\nu)) 
\label{eq:asymptotic}
\end{align}


Appendix \ref{table:ps.imp} lists the specific expressions used to approximate $f(\Delta|\boldsymbol{T},\boldsymbol{X},\boldsymbol{Y},\nu)$ for BPSA performed with caliper matching, NN matching, IPW and DR estimation.


When using asymptotic distributions to approximate $f(\Delta|\boldsymbol{T,X,Y},\nu)$, evaluation of the marginal posterior distribution in expression (\ref{eq:post.delta}) can be construed as evaluation of the ``asymptotic posterior distribution" of $\Delta$, marginalized over design uncertainty. Note that, since the derivation of $\hat{Q}(\nu)$ and $\hat{V}(\nu)$ is not typically motivated on Bayesian grounds, they do not easily incorporate prior information on $\Delta$.  Asymptotic estimators with poor finite-sample performance might propagate similar properties into posterior estimates of $\Delta$. 



\subsection{Outline of Computational Procedure for Marginalizing over Design Uncertainty} \label{analysis:computation}

Here we outline a sequential Markov-chain Monte Carlo (MCMC) procedure for evaluating the posterior distribution of causal effects, marginalized over design uncertainty as defined in expression (10), building off of previous literature \cite{kaplan,alvarez,cefalu,Spertus}. 

\subsubsection{Drawing from the posterior predictive distribution of $\nu$} \label{analysis:computation:design}

In the first stage of BPSA, multiple draws are taken from $f(\nu|\boldsymbol{T},\boldsymbol{X})$ using the following procedure:

\begin{enumerate}

\item Obtain a sample of K draws from the posterior distribution of the parameters of the propensity score model, $\alpha$.  This can be accomplished with standard MCMC routines, for example, with the R package \verb|MCMCpack|.

\item For each of the K draws from posterior distribution of $\alpha$, calculate propensity scores for each observation with the use of covariate information $\boldsymbol{X}$. Steps 1 and 2 in combination may be conceived as taking $K$ samples from the posterior predictive distribution of propensity scores.

\item For deterministic PS implementations, use the K draws from the posterior predictive distribution of propensity scores to perform K propensity score implementations, each representing a design stage $\nu_k$, indexed by $k=1,2,...,K$. For stochastic PS implementations, perform the implementation R times conditional on each set of PS for all $K$ draws, resulting in $R \times K$ values of $\nu$. 
\end{enumerate}

The output of the first stage (the design module), which may be performed in its entirety independent of outcome information, produces either $K$ or $R \times K$ draws of $\nu$ from its marginal distribution, $f(\nu|T,X)$, each representing a different design. For the analysis stage, we simplify notation and use $\nu_k$ to describe a single simulated output from the design stage, omitting the possible dependence on $R$ (in the case of probabilistic designs).  If the PS distribution is explored well enough ($K$ is large), $R$ does not need to be too large (and may even be set to 1) as the MCMC algorithm re-visits the same sets of PS multiple times and the variability of $f(\nu|\boldsymbol{T,X},\alpha)$ is explored in that way.


\subsubsection{Drawing from the conditional posterior distribution of $\Delta$} \label{analysis:computation:analysis}

Estimation in the analysis stage is conducted conditional on each of the simulated values of $\nu_k$ from the design stage. Specifically, for each value of $\nu_k, k=1,2,\ldots,K$:
\begin{enumerate}\setcounter{enumi}{3}
\item Draw $S$ samples from the conditional posterior distribution $f(\Delta|\boldsymbol{T},\boldsymbol{X}, \boldsymbol{Y}, \nu_k)$
	\begin{enumerate}[label=4.{\arabic*}a.]
	\item[] {\bf When the analysis stage entails a parametric distribution for $f(\theta|\boldsymbol{T},\boldsymbol{X},\boldsymbol{Y},\nu_k)$}
	\item Take $S$ draws from the posterior distribution of $\theta = [\beta,\sigma^2]$, a vector of parameters with the posterior distribution $f(\theta|T,X,Y,\nu) \propto L(Y|T,X,\theta,\nu)\pi(\theta)$. Let $\theta_{sk}$ represent one such draw, with $s = 1 ... S$. Again, this may be accomplished with standard MCMC procedure and a packages such as \verb|MCMCpack|.
	\item Each draw of $\theta_{sk}$  may be transformed into $\Delta_{sk}$ (ex: for stratification, $\Delta$ is a linear combination of coefficients of a regression fit where treatment assignment is interacted with strata membership). 
	\end{enumerate}
	\begin{enumerate}[label=4.{\arabic*}b.]
	\item[] {\bf When an asymptotic approximation is used for the distribution of $f(\Delta|\boldsymbol{T},\boldsymbol{X},\boldsymbol{Y},\nu_k)$}
	\item Calculate $\hat{Q}(\nu_k)$ and $\hat{V}(\nu_k)$. 
	\item Draw $S$ samples from $N(\hat{Q}(\nu_k),\hat{V}(\nu_k))$ of $\Delta_{sk}$, $s = 1 ... S$.
	\end{enumerate}
\end{enumerate}

With either step 4a or 4b of the analysis, the end result will be $K\times S$ draws from the posterior distribution $f(\Delta|\boldsymbol{T},\boldsymbol{X},\boldsymbol{Y})$, which is marginalized over design uncertainty.  If interest lies primarily in estimating the posterior mean and variance of $\Delta$, instead of retaining $S$ draws from $f(\Delta|\boldsymbol{T},\boldsymbol{X},\boldsymbol{Y},\nu_k)$ for each $k$, it is only necessary to save $\Delta_k = E(\Delta|\boldsymbol{T},\boldsymbol{X},\boldsymbol{Y},\nu_k)$ and $\sigma^2_k = Var(\Delta|\boldsymbol{T},\boldsymbol{X},\boldsymbol{Y},\nu_k)$, which can be used  with Rubin's combining rules \cite{little}, to estimate the posterior mean and variance of $\Delta$.




\subsection{Between and Within Design Uncertainty} \label{diagnose}
 The apparent parallel between the sequential modularized procedure in Section \ref{analysis:computation} and multiple imputation provides a useful analog that relates design uncertainty and analysis uncertainty to notions of ``between'' and ``within'' imputation variance.  Construing $\nu$ as a missing quantity, estimating the mean and variance of $\Delta$ with Rubin's combining rules as  $E(\Delta|\boldsymbol{T,X,Y}) = E(E(\Delta|\boldsymbol{T,X,Y}, \nu)) \approx \bar{\Delta}_K =  \frac{\sum_{k=1}^K \Delta_k}{K}$ and: 


\begin{align}
    Var(\Delta|\boldsymbol{T,X,Y}) = E(Var(\Delta | \boldsymbol{T,X,Y},\nu)) + Var(E(\Delta|\boldsymbol{T,X,Y},\nu)) \\
 \approx \bar{\sigma^2}_K + \big(1 + \frac{1}{K}\big) B^2_K = \frac{\sum_{k=1}^K \sigma^2_k}{K} +  \big(1 + \frac{1}{K}\big) \frac{\sum_{k=1}^K (\Delta_k - \bar{\Delta}_K)^2 }{K - 1}
\end{align}

Where $K$, $\Delta_k$ and $\sigma^2_k$ are as defined in Section \ref{analysis:computation:analysis} and can be calculated whether a parametric distribution or asymptotic approximation is used in the analysis stage.



The quantity AEU relates to within-design variability ($\bar{\sigma^2}_K$), quantifying the variability in the distribution of $\Delta$, conditional on $\nu$. Design uncertainty (both DDU and DEU) relate to between-design variability ($B^2_K$) across the samples from the distribution of $\nu$. Plummer \cite{plummer} draws parallels between the sequential (or modularized) MCMC procedure of Section \ref{analysis:computation} and concepts in the MI literature, where the reliability of posterior inference relates to the interplay between these two elements of uncertainty. Equating $K$ to the number of ``imputed" designs, we define an expression for the proportion of total variability in $\Delta$ attributable to the between-design uncertainty, $PROP_{DU}$, which is closely related to the proportion of missing information in a multiple imputation procedure \cite{rubin_mi_1996}:



\begin{align}
    PROP_{DU} = \frac{B^2_{K}}{B^2_{K} + \bar{\sigma}^2_{K}}
\end{align}




In MI, when the percentage of missing information is high, inference on $\Delta$ must be treated with caution. The same principle applies to design uncertainty and the relative influence of between-design variability as quantified in $PROP_{DU}$. Section \ref{sim} examines the relationship between $PROP_{DU}$ and BPSA performance in finite samples across several scenarios with varying amounts of design uncertainty.  Examining BPSA through this lens also points towards expected behavior in large samples.  As $n$ approaches $\infty$, the posterior distribution of the propensity score model parameters ($f(\alpha|\mathbf{X,T})$) will converge to point mass at the MLE. For implementations with no DDU (i.e., point mass for $f(\nu|\mathbf{X,T},\alpha)$), this would result in between-design variability approaching 0 and the within-design variance concentrating around the asymptotic $Var(\Delta|\mathbf{X,Y,T},\nu)$, which, in combination, would result in BPSA estimators exhibiting the same asymptotic properties (e.g., coverage) as their Frequentist PSA counterparts.

\section{Simulation Study to Assess Design and Analysis Uncertainty under varying PS models and implementations} \label{sim}


Here we provide a simulation study to evaluate how different ``design choices'' made in PSA impact design and analysis uncertainty.  Specifically, we implement BPSA to examine design and analysis uncertainty under the different PS implementations from Section \ref{ps.design:imp} and across PS models that include various types of covariates (i.e., confounders, instrumental, prognostic, and noise variables). To focus on the impact of design choices, the different PS models and implementations will be considered in Monte Carlo replicates from a single data generating mechanism.  

While the primary goal is to examine the different components of uncertainty within BPSA and how quantities such as $PROP_{DU}$ relate to performance,  standard (i.e., Frequentist) PSA estimators will be included for comparison, without claims as to the superiority of the Bayesian vs. Frequentist approach.  


\subsection{Data generating mechanism}
200 replicated data sets are simulated to have $n=1000$ observations and 20 uncorrelated normally-distributed covariates with mean 0 and variance 1. 5 of these covariates are true confounders associated with both $T$ and $Y$, while 5 are instruments (associated with $T$ only), 5 are prognostic (associated with $Y$ only) and 5 are noise (associated with neither $T$ nor $Y$) variables. Treatment assignment $T$ is simulated from a Bernoulli distribution with probability of treatment specified by the logistic regression described in Section \ref{ps.design:deu}, with intercept set to zero and coefficients $\alpha$ set to 0.75 for all confounders and instrumental variables and 0 for all prognostic and noise variables. Outcomes, $Y$, are simulated from a linear regression model of the form $Y = 1 + \Delta T + \beta \mathbf{X} + \epsilon$, with $\Delta$ set to 1.5, and $\beta$ set to [0.1, 0.2, 0.3,0.4,0.5] for the five confounders, 0.5 for all prognostic variables and 0 for noise and instrumental variables. $\epsilon$ is a normally-distributed random error with mean 0 and variance 1.


\subsection{BPSA and PSA procedures} \label{sim:bpsa}
We analyzed simulated datasets with BPSA (procedure described in Section \ref{analysis:computation} with $K = 1000$, $S = 200$ and $R = 1$ for caliper matching) and standard PSA (which conditions analysis on the MLE of the PS) in the context of four different PS model specifications and six PS implementations. 

The four PS model specifications were logistic regressions including the following sets of variables:

\begin{enumerate}
    \item Only confounders
    \item Confounders and instrumental variables
    \item Confounders, instrumental and prognostic variables
    \item Confounders, instrumental, prognostic and noise variables
\end{enumerate}

Note that all scenarios correctly specify the functional form of the PS model and all include at least the confounders necessary to satisfy the ignorability assumption. When calculating the DR estimator, the potential outcome model utilizes the same covariates as the PS model. 


With each of the two PS model specifications, we employ the following PS implementations: 1-1 caliper matching with a caliper of 0.5, 1-5 caliper matching with a caliper of 0.5, 1-1 NN matching with a caliper of 0.5, quintile stratification, and weighting with IPW and DR estimators. Different ratios used for caliper matching were designed to influence the amount of DDU in the implementation; increasing the number of control matches decreases the amount of DDU.

Under BPSA, the analysis stage following stratification is performed with parametric modeling of the conditional distribution of $\Delta$ (see Section \ref{analysis:parametric}) while the analysis stage following all other implementations involve asymptotic approximation of the conditional distribution of $\Delta$ (see Section \ref{analysis:asymptotic}). 

Under PSA, the analysis stage after all implementations utilizes asymptotic estimators for both the treatment effect and variance, where ``robust" variance estimators are employed after caliper matching (with both ratios), NN matching and IPW estimation and a standard OLS regression is performed for stratification utilizing the same linear model as BPSA (which interacts treatment and strata membership). The variance estimator for DR under PSA is the conditional variance estimator for $\Delta$ which is utilized in BPSA. Details of all treatment effect and variance estimators may be found on Appendix \ref{table:ps.imp}.

\subsection{Simulation Results}


\begin{figure}[t]
\centerline{\includegraphics[width=\textwidth]{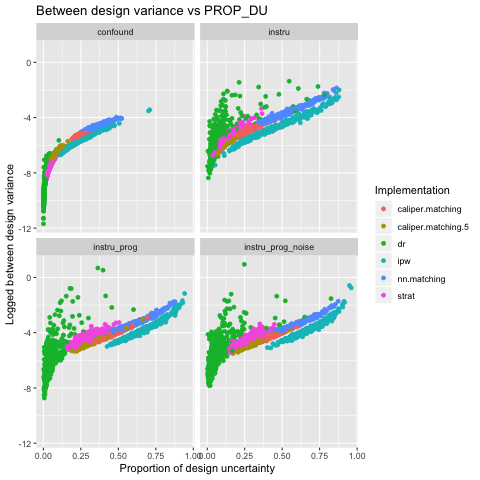}}\caption{Relationship between between-design variability and $PROP_{DU}$ under different PS models and PS implementations \label{fig:sim.bv}}
\end{figure}

\begin{figure}[t]
\centerline{\includegraphics[width=\textwidth]{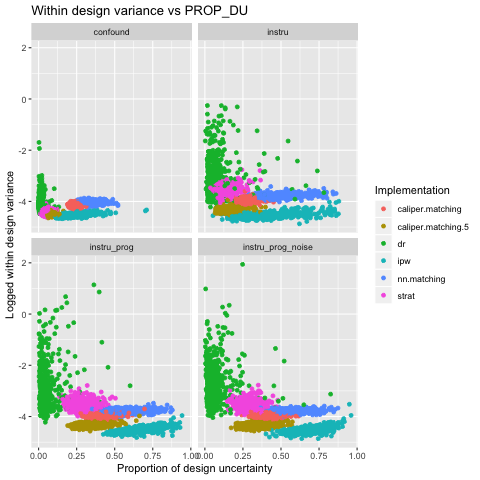}}\caption{Relationship between within-design variability and $PROP_{DU}$ under different PS models and PS implementations \label{fig:sim.wv}}
\end{figure}

\begin{figure}[t]
\centerline{\includegraphics[width=\textwidth]{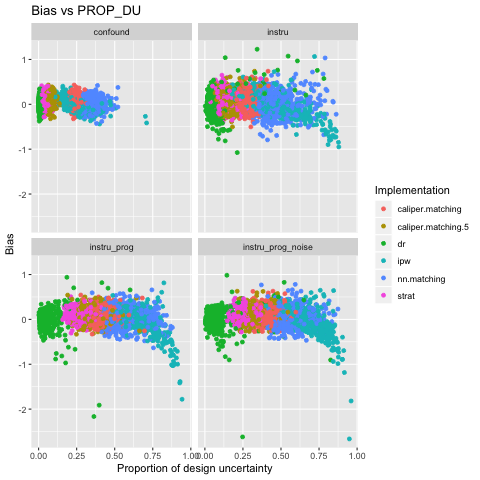}}\caption{Relationship between bias and $PROP_{DU}$ under different PS models and PS implementations \label{fig:sim.bias}}
\end{figure}

A complete table of simulation results, including bias, empirical variance, and mean squared error (MSE) of posterior mean estimates of $\Delta$ across replications, average between-design, within-design and total estimated variance across replications, average $PROP_{DU}$ across replications, and coverage of 95\% posterior intervals for the 24 combinations of PS model and implementation can be found in Table \ref{table:sim}. Plots depicting how between design variance (log transformed), within design variance (log transformed) and bias vary across levels of $PROP_{DU}$ appear in Figures \ref{fig:sim.bv}, \ref{fig:sim.wv}, and  \ref{fig:sim.bias}, respectively.



Compared to a PS model that includes only the confounders, adding additional instrumental and prognostic variables increases average between-design variability for all implementations (Figure \ref{fig:sim.bv}), with less pronounced impact on within-design variability (Figure \ref{fig:sim.wv}). As a consequence, adding these variables to the PS model leads to higher average values of $PROP_{DU}$ within all implementations. 



Under all PS models, the implementations display clear patterns in the amount of $PROP_{DU}$ they exhibit. DR displays the lowest $PROP_{DU}$ under all models, followed by stratification and 1-5 caliper matching. This aligns with the expectation that these implementations are relatively insensitive to moderate perturbations in the PS. On the other hand, IPW and NN matching display the highest values in $PROP_{DU}$, indicating that even mild perturbations in the PS may result in very different treatment effect estimates. 



Estimates of $\Delta$ using IPW display the clearest trend in performance, where high $PROP_{DU}$ is associated with more erratic effect estimates in terms of large bias (Figure \ref{fig:sim.bias}), poor coverage and high variance across replications (Table \ref{table:sim}). For IPW, adding instrument and prognostic variables has the most impact on the between-design variance of IPW estimates (Figure \ref{fig:sim.bv}), while within-design variance is small and largely unaffected by changes to the PS model specification (Figure \ref{fig:sim.wv}), resulting in large differences in average $PROP_{DU}$ across PS models. Performance of $\Delta$ estimates is consistent with the expectation that adding many covariates into the PS model can result causal estimates that are sensitive to extreme weights and low covariate overlap\citep{robins_marginal_2000}.



The performance of NN matching, on the other hand, does not suffer as substantially as IPW under PS models which generate high $PROP_{DU}$. Under PS models with instrumental, prognostic and noise variables, NN matching estimates $\Delta$ with less empirical variability across replications than IPW (Table \ref{table:sim}) under all PS models since frequency weights are less likely to take on extreme values compared to IP weights. Coverage levels for NN matching are closer to nominal than IPW under PS models 2-4 due to the comparatively higher average total variance. This is primarily due to NN matching's between-design variability (Table \ref{table:sim}) which increases as more covariates (regardless of type) are added to the PS model, created by averaging estimates over designs that consist of different subsets of the data (on average 71\% under the confounders-only PS model and on average 63\% under other PS models).




Both caliper matching implementations show similar trends to NN matching in terms of the relationship between $PROP_{DU}$ and $\Delta$'s estimation performance, but with lower average between-design variability, empirical variability, and within-design variability within each PS model. Comparing NN matching and 1-1 caliper matching, 1-1 caliper matching displays lower between-design variability because the pool of possible matches displays few changes across draws from the distribution of PS due to the generous caliper (Section \ref{ps.design:imp:caliper}), whereas the ``nearest neighbor" of a given observation may change with every PS draw, resulting in greater differences between designs. 1-5 caliper matching displays less between-design variability than 1-1 caliper matching since the average of 5 matches naturally has less variability across designs than a single match, an example of decreasing DDU. Among all the matching procedures, 1-5 caliper matching shows the lowest within-design variability (and tends to include the highest proportion of the observed data - 92\% on average - in the design), and 1-1 caliper matching exhibits slightly lower within-design variability than NN matching. Both caliper matching procedures are similar to NN matching in their relative insensitivity to $PROP_{DU}$ in terms of within-design variability. Importantly, the additional DDU associated with the caliper matching does not translate into increased between-design variability in estimates of $\Delta$, relative to NN matching (which exhibits no DDU).

Stratification exhibits the lowest average between-design variability among virtually all implementations and across all PS models, as observations generally remain in the same PS strata across multiple draws from the posterior distribution of the PS (Table \ref{table:sim}).  Accordingly, stratification displays low $PROP_{DU}$ across all PS models, and lower empirical variability estimates of $\Delta$ than the matching or IPW procedures.  The stratification approach exhibits slightly more bias than the other procedures, likely due to the relatively small number of strata.


The DR estimator is unique among our considered implementations as it is the only one which incorporates an explicit model for potential outcomes that includes direct covariate adjustment. Thus, it is not surprising that its performance deviates from the general trend of the other implementations. First, within each PS model most replications under DR exhibit much lower $PROP_{DU}$ compared to other implementations. While most implementations display only small differences in within-design variability across PS models, DR displays large increases in average within-design variability in response to adding instrumental, then prognostic variables to the PS model. For most of the PS models considered, DR displays the largest within-design variability on average. Between-design variability, however, appears to only moderately increase as instrumental and prognostic variables are added to the PS model, explaining DR's relatively smaller $PROP_{DU}$ under all PS models. The inclusion of the outcome model in the DR estimate leads to the comparatively low between-design variability, as perturbations in the propensity score have less impact in effect estimation than the (correctly specified) outcome model.  




With the exception of DR, empirical variance of posterior mean estimates of $\Delta$ across replications demonstrate a positive relationship with $PROP_{DU}$ (Table \ref{table:sim}) within each PS model, indicating a correspondence between between-design variability and variability of point estimates in repeated samples. 
Designs (combinations of PS models and implementations) which exhibit high $PROP_{DU}$ also exhibit more variability in treatment effect estimates across replication and thus require larger estimated variances in order to achieve nominal coverage. 


Results of a standard PSA analysis are presented in Table \ref{table:psa}. For the three matching implementations and stratification, BPSA results in lower empirical variability  and MSE due to the averaging over multiple possible matched or stratified designs. This trend is more evident under PS models which include more than the necessary confounders. In contrast, BPSA versions of IPW and DR exhibit slightly higher empirical variability and MSE than PSA, with this difference also most pronounced in the PS models that include prognostic, instrument, and/or noise variables.  This may be a consequence of extreme weights, where averaging over multiple designs includes those with weights extreme enough to provide erratic estimates and increased variability.

Standard PSA under the three matching implementations and IPW achieve nearer to nominal coverage than BPSA under most PS models. For DR, PSA performs similarly to BPSA under all PS models except for the PS model with confounders and instruments, where BPSA achieves nominal coverage while PSA intervals do not. Finally, with stratification both PSA performs similarly to BPSA with regard to coverage under all PS models. 

\section{Investigating the Effect of High Power Plant Emissions on Ambient Pollution} \label{application}

An ongoing analysis in Cummiskey et. al. \cite{cummiskey} is evaluating the association between long-term exposure to emissions from coal-fired power plants and Ischemic Heart Disease (IHD) hospitalizations among Medicare beneficiaries. One important feature of this analysis is the way in which the analysis incorporates ambient fine particulate matter (particles less than 2.5 micrometers in diameter, denoted PM).  Ambient PM concentrations are expected to be derived in part from coal-fired power plant emissions in some regions, and PM is known to be associated with a variety of adverse health outcomes \cite{pope}.  While the primary analysis in Cummiskey et. al.\cite{cummiskey} investigates the link between coal emissions and IHD without adjusting for PM, a secondary analysis regards PM as an adjustment covariate and briefly evaluates whether PM could be ruled out as a possible mediator of the relationship between coal emissions and IHD hospitalizations.   Here we revisit the secondary analysis of Cummiskey et. al. \cite{cummiskey}, deploying the methods described in Section \ref{sim:bpsa} to investigate the extent to which a binary metric of high/low coal emissions exposure causally impacts annual average PM. Details in Cummiskey et. al. \cite{cummiskey} describe the creation of a binary metric of coal power plant exposure derived using a reduced-complexity chemical transport model (InMap) \cite{tessum2017inmap}, as well as the data-fusion derived estimates of PM \cite{di2016assessing}. 

We examine the effect of elevated coal emissions on ambient particle concentrations measured at  22,723 zip codes in the Northeast, Southwest and Industrial Midwest regions of the United States. The analysis adjusts for 17 possible confounders, including location of zip code (latitude and longitude), climate (temperature and humidity), population density, demographics (e.g. racial makeup, education, household income, gender) as well as the makeup of the residential areas in each zip code (e.g. urban/rural, real estate value). A full list of covariates considered can be found in Table \ref{table:app.description}. Further details on consolidating the zip-code-level covariates (retrieved from 2000 US Census data) may be found in Henneman et. al \cite{cummiskey}. 

\begin{figure}[t]
\centerline{\includegraphics[width=0.7\textwidth,height = 10cm]{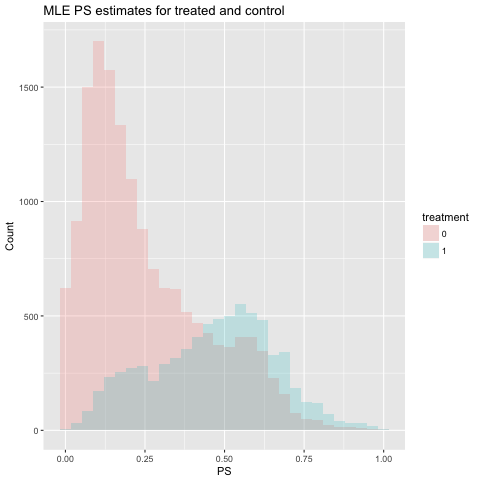}}
\caption{Histogram of estimated propensity scores under treated and control observations \label{fig:app.hist}}
\end{figure}

Among the 22,723 observations in this dataset, 7,211 are classified as ``exposed" (to high coal emissions) and 15,512 are control observations. Figure \ref{fig:app.hist} visualizes the overlap of estimated propensity scores among treated and control observations using maximum likelihood estimates of the PS, and Table \ref{table:app.description} contains average values of included confounders compared across exposure status. On average, control zip codes are located further West, have lower population densities, are less urban and more rural, and exhibit a lower median household income and higher percent poverty. 

Analysis was performed with the PSA and BPSA procedures described in Section \ref{sim:bpsa}, compared within the contexts of quintile stratification, caliper matching with replacement, NN matching with replacement (both with caliper = 0.5 and a 1-1 ratio), and weighting with IPW and DR estimators. In this analysis, we focus on estimating the ATE, but employ caliper and NN matching methods which measure the ATT for purposes of illustration. 

\begin{figure}[t]
\centerline{\includegraphics[width=0.7\textwidth]{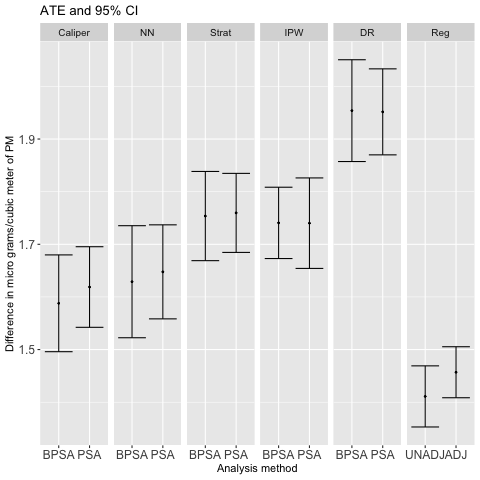}}
\caption{Estimated ATEs and constructed 95\% confidence/credible intervals created via PSA and BPSA \label{fig:app.results}}
\end{figure}

\begin{center}
\begin{table}[b]%
\centering
\caption{Application data: table of unadjusted averages of covariates across treatment
\label{table:app.description}}%
\begin{tabular*}{300pt}{@{\extracolsep\fill}lcc@{\extracolsep\fill}}%
\toprule
 & \textbf{Unexposed} & \textbf{Exposed} \\ 
 \midrule
  Number of observations & 15,512 & 7,211 \\ 
  Avg. estimated PS & 0.25 & 0.46 \\ 
  Latitude & 38.30 & 38.37 \\ 
  Longitude & -82.84 & -79.46 \\ 
  County smoking rate & 0.27 & 0.26 \\ 
  Total population & 10,791 & 13,231 \\ 
  Percent residing in rural area & 0.58 & 0.49 \\ 
  Percent of white residents & 0.85 & 0.82 \\ 
  Percent of African-American residents & 0.11 & 0.13 \\ 
  PctHighSchool & 0.35 & 0.35 \\ 
  Median household income & 39,385 & 43,594 \\ 
  Percent living below poverty threshold & 0.13 & 0.12 \\ 
  Percent female & 0.51 & 0.51 \\ 
  Percent of housing units occupied & 0.87 & 0.90 \\ 
  Percent who have moved in past 5 years  & 0.42 & 0.42 \\ 
  Median house value & 102,070 & 121,912 \\ 
  Population per Sq. Mile & 1,167 & 2,455 \\ 
  Avg. temperature (2005) & 286 & 286 \\ 
  Avg. relative humidity (2005) & 0.0086 & 0.0085 \\ 
   \bottomrule
\end{tabular*}
\end{table}
\end{center}

\begin{center}
\begin{table}[b]
\centering
\caption{Estimated treatment effects of high power plant emissions on ambient pollution \label{table:app.details}}
\begin{tabular}{lll}
                          & \multicolumn{2}{l}{\textbf{ATE {[}95\% CI{]}}} \\
\textbf{UNADJ regression} & \multicolumn{2}{l}{1.41 {[}1.35,1.47{]}}       \\
\textbf{ADJ regression}   & \multicolumn{2}{l}{1.46 {[}1.41,1.51{]}}       \\
                          & \textbf{BPSA}          & \textbf{PSA}          \\

\textbf{NN matching}      & 1.63 {[}1.52,1.74{]}   & 1.65 {[}1.56,1.74{]}  \\
\textbf{Caliper matching} & 1.59 {[}1.50,1.68{]}   & 1.63 {[}1.55,1.70{]}  \\
\textbf{Stratification}   & 1.75 {[}1.67,1.84{]}   & 1.76 {[}1.68,1.83{]}  \\
\textbf{IPW}              & 1.74 {[}1.64,1.85{]}   & 1.74 {[}1.65,1.83{]}  \\
\textbf{DR}               & 1.95 {[}1.86,2.05{]}   & 1.95 {[}1.87,2.03{]} 
\end{tabular}
\end{table}
\end{center}

\begin{center}
\begin{table}[b]%
\centering
\caption{Between- and within-design variance under BPSA
\label{table:app.vars}}%
\begin{tabular*}{300pt}{lllll}%
\toprule
                                                                         & \textbf{\begin{tabular}[c]{@{}l@{}}Total\\ variance\end{tabular}} & \textbf{\begin{tabular}[c]{@{}l@{}}Between-\\ design\\ variance\end{tabular}} & \textbf{\begin{tabular}[c]{@{}l@{}}Within-\\ design\\ variance\end{tabular}} & \textbf{$PROP_{DU}$} \\
                                                                         \midrule
\textbf{NN matching}                                                     & 0.0029                                                            & 0.0018                                                                        & 0.0012                                                                       & 0.59                 \\
\textbf{\begin{tabular}[c]{@{}l@{}}1-1 Caliper \\ matching\end{tabular}} & 0.0022                                                            & 0.0011                                                                        & 0.0011                                                                       & 0.49                 \\
\textbf{DR}                                                              & 0.0024                                                            & 0.0003                                                                        & 0.0020                                                                       & 0.11                 \\
\textbf{IPW}                                                             & 0.0012                                                            & 0.0005                                                                        & 0.0006                                                                       & 0.46                 \\
\textbf{Stratification}                                                  & 0.0019                                                            & 0.0002                                                                        & 0.0016                                                                       & 0.13\\        \bottomrule        
\end{tabular*}
\end{table}
\end{center}

Estimated treatment effects, 95\% uncertainty intervals and analysis details may be found in Table \ref{table:app.details}, and are depicted graphically in Figure \ref{fig:app.results}. Presented on the same scale for comparison are also estimates and intervals constructed via fully adjusted and unadjusted linear regressions. Between-, within- and total variance for all implementations utilized with BPSA may be found in Table \ref{table:app.vars}. NN matching, 1-1 caliper matching and IPW all display high $PROP_{DU}$ at 0.59, 0.49 and 0.46, respectively, while DR and stratification display relatively lower $PROP_{DU}$, at 0.11 and 0.13, respectively (Table \ref{table:app.vars}). This mirrors our simulation findings, as DR and stratification display less between-design variability compared to the matching and IPW methods.

Both the fully adjusted and unadjusted linear regressions estimate lower treatment effects than any of the propensity score analysis methods. IPW and stratification produced similar point estimates for the average treatment effect ($\approx 1.75$) while both matching implementations display a lower estimation of the ATT. The DR estimator displays the highest ATE estimate, and its confidence intervals do not overlap with that of any other implementation. The discrepancy between the DR estimate and the IPW estimate may arise due to the relatively high $PROP_{DU}$, to which IPW estimator was shown (in the simulation) to be sensitive.  The discrepancy with the stratification estimator may relate to the relatively small number of strata used.


While point estimates of treatment effects are nearly identical between PSA and BPSA procedures in the context of DR, IPW and stratification, BPSA point estimates are lower than PSA point estimates in the context of NN matching and caliper matching. This may be attributed to the large pool of control observations available compared to treated observations. Both matching algorithms on average use only 50\% of the available observations in their treatment effect estimation, and selection of controls into the matched set is sensitive to design uncertainty arising from both DEU (the PS model includes many possible confounders, some of which may be instrumental variables) as well as DDU (in the case of caliper matching). By averaging over multiple matched sets, BPSA was shown to produce more stable estimates of the treatment effect in Section \ref{sim} and it is reflected in this analysis via the contrast between PSA and BPSA-estimated treatment effects.

As witnessed in the simulation study, BPSA produces intervals similar in width to PSA when performed with stratification and DR. Only under IPW does BPSA create narrower intervals than PSA, a result which occurred in the simulation study when non-confounders were included in the PS model. The most unexpected result occurs under 1-1 NN and caliper matching, where BPSA produces slightly wider intervals than PSA, a result not observed in the simulation study. This is due to the difference in proportion of treated observations, which was on average equal to the proportion of control observations in the simulated data but much lower than the proportion of control observations in the power plant data. The high number of control observations creates variability in the treatment effect estimate across draws from the PS distribution, as reflected in the high between-design variance and $PROP_{DU}$ displayed by both matching methods (Table \ref{table:app.vars}). The design uncertainty captured by the between-design variability translates into a higher total variance than PSA.  



Overall, the results of the analysis provide the consistent message that elevated exposure to coal-fired power plant emissions causally increases the overall annual concentration of ambient fine particulate matter, underscoring the care with which the secondary analysis in Henneman et. al\cite{cummiskey} should be interpreted.


\section{Discussion}\label{discussion}


This paper formalizes a sequential Bayesian procedure for marginalizing causal effect estimates over uncertainty associated with the design stage of a PSA.  The ability to distinctly define design and analysis uncertainty aids understanding of the sensitivity of causal effect estimate performance to decisions made during the approximation of the design of a randomized study with propensity scores. The procedure outlined in this paper synthesizes a variety of related ideas that have recently appeared under the topic of Bayesian methods for propensity score analysis \cite{mccandless,alvarez,kaplan,jacob_better_2017,plummer,zigler_model_2013}. 





The simulation study explores how design decisions such as covariate selection for the PS model and choice of PS implementation dictate not only the quantity of design uncertainty which results, but how it is propagated into estimates of causal effects. In particular, the simulation showed how a summary measure such as $PROP_{DU}$ quantifying the relative amount of ``between'' and ``within'' design uncertainty can be useful for understanding the impact of design uncertainty on posterior variability in estimates of $\Delta$. In general, higher $PROP_{DU}$ was associated with higher variability of $\Delta$ point estimates across replications, though specific impacts of this with regard to coverage and MSE varied between implementations. The primary goal of the simulation section was not to establish an superiority to BPSA relative to standard PSA. While BPSA resulted in lower MSE for the matching implementations and stratification and comparable coverage to PSA under DR and stratification, neither method appeared uniformly superior. In order to focus on the nuances around between- and within- design uncertainty, the simulation study considered the implications of design choices within a single data generation, further limiting our ability to make general claims of the relative performance of BPSA vs. standard PSA approaches.

The work here shares conceptual similarities to that of Branson \cite{branson}, who compares the analysis stage of a PSA to that of a RCT with randomization restricted to pre-determined bounds on covariate balance. However, while Branson describes a conditional distribution of treatment effects which accounts for certain limitations placed on PS implementations, the paper does not address uncertainty arising from models or implementations utilized in the design stage nor how to propagate it into estimation of the treatment effect. Furthermore, Branson primarily focuses on PS matching while this paper extends to any PS implementation. Our discussion also construes design uncertainty broadly and does not detail all the possible factors of design uncertainty. However, this subject matter may be related to topics such as low covariate overlap across treatments, low treatment prevalence and highly correlated confounders. For example, IPW is known to give erratic finite-sample performance in the presence of low covariate overlap, which we have replicated in our simulations \cite{fanli}. Work exploring these connections in further detail would be a promising extension. Other extensions could incorporate non-parametric, machine-learning \cite{lee} or flexible \cite{zandd} PS estimation methods which have become recently popular. 

We framed BPSA as a special instance of modularization in Bayesian inference, where the mechanics of posterior updating were seperated in the ``design'' module and the ``analysis'' module.  The overlap between posterior inference on modules and multiple imputation is discussed in Plummer et. al. \cite{plummer}, and this paper makes ample use of this connection to frame marginalizing over design uncertainty through the lens of between- and with- design uncertainty, as would be relevant to an analysis of missing data.   Despite the usefulness of the parallel between BPSA and multiple imputation, better understanding of the limits of this commonality could further improve understainding. For instance, $\nu$ in the present framework is an entirely synthetic parameter more akin to a latent variable than a missing quantity with a probabilistic missing-data response mechanism, which is key to establishing both Bayesian and Frequentist validity for estimators based on multiple imputation.  Further exploration of the role that the theory of multiple imputation can play in this context is an interesting direction for future work.

Though propensity score analysis is widely used in various fields including epidemiology and econometrics, notions of design uncertainty are rarely considered explicitly. Bayesian propensity score analysis provides a versatile computational procedure which marginalizes over all components of design uncertainty in a clearly defined manner. Though the performance of BPSA varies across implementations, it has been regarded as an attractive alternative to conditioning inference on a single estimate of the propensity score \cite{alvarez}. Explication and exploration of the ideas presented here can hopefully ground future work investigating how Bayesian methods can add to the existing literature on propensity score analysis.

\section*{acknowledgements}
Data utilized in the analysis of power plant emissions are available from Drs. Joel Schwartz and Qian Di. Restrictions apply to the availability of these data, which were used with permission for this study. Simulation data, a simulated set of power plant emission data and code to implement the simulation study and power plant analysis are available at: https://github.com/shirleyxliao/Uncertainty-in-the-Design-Stage-of-Two-Stage-Bayesian-Propensity-Score-Analysis

We thank Dr. Kevin Cummisky for aiding the analysis of the power plant emissions data, Drs. Joel Schwartz and Qian Di for providing the data-fused air pollution predictions, and Drs. Francesca Dominici and Sebastien Haneuse for helpful feedback and discussion. This work was supported by research funding from NIHR01GM111339, NIHR01ES026217, EPA 83587201, and HEI 4953. Its contents are solely the responsibility of the grantee and do not necessarily represent the official views of the USEPA. Further, USEPA does not endorse the purchase of any commercial products or services mentioned in the publication.


\bibliography{sample}

\appendix

\section{Model specification for examined implementations} \label{table:ps.imp}
\subsection{Stratification}
Let $\nu_i \in [1,...,S]$ represent the strata which observation $i$ is assigned to. We model the outcome with a linear regression: 

\begin{align}
E(Y_i) = \beta_0 + \beta_1 T_i + \sum_{s=2}^S \big[ \beta_{2s} \mathbb{1}_{(\nu_i=s)} + \beta_{3s} T_i \mathbb{1}_{(\nu_i=s)} \big]
\end{align}

Let $P_{st}$ represent the proportion of those in treatment group $t$ who were assigned to stratum $s$, thus $P_{st} = \frac{\sum_{i=1}^n \mathbb{1}_{(\nu_i = s)}\mathbb{1}_{(T_i = t)}}{\sum_{i=1}^n \mathbb{1}_{(T_i = t)}}$ Our estimate of $\Delta$ is then a linear combination of $\beta$s: 

\begin{align}
\Delta = \frac{\sum_{i=1}^n Y_i T_i}{\sum_{i=1}^n T_i} - \frac{\sum_{i=1}^n Y_i (1-T_i)}{\sum_{i=1}^n (1-T_i)}  = \beta_1 + \sum_{s=2}^S \big[ \big(P_{s1} - P_{s0}\big)\beta_{2s} + P_{s1} \beta_{3s} \big]
\end{align}

Where the effect difference in level $s=1$ is represented by $\beta_1$. In Sections \ref{sim} and \ref{application}, quintile stratification is utilized, setting $S = 5$. 
When performing standard Frequentist PSA, the MLE estimate for $\boldsymbol{\hat{\beta}}$ may be substituted for $\boldsymbol{\beta}$ to calculate $\Delta$. The asymptotic form for $Var(\Delta)$ was derived by Lunceford and Davidian \cite{lunceford}, which is functionally equivalent to performing a transformation of the variance-covariance matrix for $\boldsymbol{\beta}$. We calculated the latter when performing PSA in Section \ref{sim}. 

Since a parametric likelihood may be recovered from this specification of the outcome model, when performing BPSA $E(\Delta|\boldsymbol{T,X,Y,\nu})$ and $Var(\Delta|\boldsymbol{T,X,Y,\nu})$ are simply the conditional posterior mean and variance of $\Delta$, respectively, as outlined in Section \ref{analysis:parametric}. In practice, this involves performing a Bayesian linear regression conditional on strata assignments and obtaining the posterior mean and variance of multiple draws. We utilized a flat prior on $\boldsymbol{\beta}$. 

\subsection{DR}\label{apx:dr}
Doubly-robust effect estimation is performed as described in Funk et. al. \cite{funk}. Let $\hat{Y}_1$ represent counterfactual outcomes for an all-treated sample, estimated with the following model:

\begin{align}
E(Y_i | T_i = 1) = \beta_0 + \sum_{q=1}^p \beta_{q} X_{iq}
\end{align}

$\boldsymbol{\hat{\beta}}$ is estimated by fitting this regression on the subsample of treated observations, then $\hat{Y}_1$ is predicted for all observations. With a similar process, $\hat{Y}_0$ is also predicted. The following formula is then utilized to estimate $E(\hat{\Delta}^*(\nu))$, where $\nu_i$ is defined in Equation \ref{eq:weights}:

\begin{align}
    \big[\frac{1}{n}\sum_{i=1}^n \frac{T_i Y_i}{\hat{e}_i} - \frac{\hat{Y_1}(T_i - \hat{e}_i)}{\hat{e}_i}\big] - \big[\frac{1}{n}\sum_{i=1}^n \frac{(1-T_i) Y_i}{(1-\hat{e}_i)} - \frac{\hat{Y_0}(T_i - \hat{e}_i)}{1 - \hat{e}_i}\big] \\
    = \big[\frac{1}{n}\sum_{i=1}^n \nu_i (Y_i - \hat{Y}_1) + \hat{Y}_1 \big] - \big[\frac{1}{n}\sum_{i=1}^n \nu_i (Y_i - \hat{Y}_0) + \hat{Y}_0 \big]
\end{align}

While the following formula from Lunceford and Davidian \cite{lunceford} is utilized to estimate $Var(\hat{\Delta}^*(\nu))$:

\begin{align} 
  \frac{1}{n} \Bigg[\Big[\frac{1}{n} \sum_{i=1}^n \big[\frac{T_i(Y_i-\hat{\mu}_{IPW,1})}{\hat{e}_i} - \frac{(1-T_i)(Y_i-\hat{\mu}_{IPW,0})}{1-\hat{e}_i}\big]^2 \Big] -
  \Big[\frac{1}{n} \sum_{i=1}^n \big[\sqrt{\frac{1-\hat{e}_i}{\hat{e}_i}} (\hat{Y}_1 - \hat{\mu}_{DR,1}) +  [\sqrt{\frac{\hat{e}_i}{1-\hat{e}_i}} (\hat{Y}_0 - \hat{\mu}_{DR,0})\big]^2 \Big]\Bigg] \\
  = \frac{1}{n} \Bigg[\Big[\frac{1}{n} \sum_{i=1}^n \big[\nu_iT_i(Y_i-\hat{\mu}_{IPW,1}) - \nu_i(1-T_i)(Y_i-\hat{\mu}_{IPW,0})\big]^2 \Big] -
  \Big[\frac{1}{n} \sum_{i=1}^n \big[(\nu_i-1)^{T_i - \frac{1}{2}} (\hat{Y}_1 - \hat{\mu}_{DR,1}) +  (\nu_i - 1)^{\frac{1}{2} - T_i} (\hat{Y}_0 - \hat{\mu}_{DR,0}) \big]^2 \Big]\Bigg]
\end{align}

Where $\hat{\mu}_{IPW,1} = \frac{\sum_{i=1}^n \frac{T_i Y_i}{\hat{e}_i}}{\sum_{i=1}^n \frac{T_i}{\hat{e}_i}} = \frac{\sum_{i=1}^n \nu_i T_i Y_i}{\sum_{i=1}^n \nu_i T_i}$, $\hat{\mu}_{IPW,0} = \frac{\sum_{i=1}^n \frac{(1-T_i) Y_i}{1-\hat{e}_i}}{\sum_{i=1}^n \frac{1-T_i}{1-\hat{e}_i}} = \frac{\sum_{i=1}^n \nu_i (1-T_i) Y_i}{\sum_{i=1}^n (1-T_i)\nu_i}$, $\hat{\mu}_{DR,1} = \frac{1}{n}\sum_{i=1}^n \frac{T_i Y_i}{\hat{e}_i} - \frac{\hat{Y_1}(T_i - \hat{e}_i)}{\hat{e}_i} = \frac{1}{n}\sum_{i=1}^n \nu_i(Y_i - \hat{Y}_1) + \hat{Y}_1$ and $\hat{\mu}_{DR,0} = \frac{1}{n}\sum_{i=1}^n \frac{(1-T_i) Y_i}{(1-\hat{e}_i)} - \frac{\hat{Y}_0(T_i - \hat{e}_i)}{1 - \hat{e}_i} = \frac{1}{n}\sum_{i=1}^n \nu_i (Y_i -\hat{Y}_0) + \hat{Y}_0$. 

\subsection{IPW} \label{apx:ipw}
When utilizing IPW with BPSA, the conditional distribution of $\Delta$ is estimated asymptotically. 

The following formula is used to estimate $E(\hat{\Delta}^*(\nu))$, where $\nu_i$ is defined in Equation \ref{eq:weights}:

\begin{align}
    \Delta = \frac{\sum_{i=1}^n \frac{T_i Y_i}{\hat{e}_i}}{\sum_{i=1}^n \frac{T_i}{\hat{e}_i}}  - \frac{\sum_{i=1}^n \frac{(1-T_i) Y_i}{1-\hat{e}_i}}{\sum_{i=1}^n \frac{1-T_i}{1-\hat{e}_i}} \\
    = \frac{\sum_{i=1}^n \nu_i T_i Y_i}{\sum_{i=1}^n \nu_i T_i} - \frac{\sum_{i=1}^n \nu_i (1-T_i) Y_i}{\sum_{i=1}^n (1-T_i)\nu_i}
\end{align}

This is equivalent to the estimation of $\hat{\beta}_1$ in the following regression, when weights are set equal to $\nu_i$:

\begin{align}\label{eq:msm}
    E(Y_i) = \beta_0 + \beta_1 T_i
\end{align}


For PSA, $\hat{Var}(\hat{\Delta})$ is estimated with the Hubert-White sandwich estimator \cite{kman}, which is known to be conservative. This variance estimator leads to double-counting of design uncertainty when used to estimate $\hat{V}(\nu)$ in BPSA. 

We instead estimate $Var(\Delta^*(\nu))$ with the following estimator from Lunceford and Davidian \cite{lunceford} which assumes the ``true" propensity score (and thus true $\nu_i$) is known: 

\begin{align}
   \frac{1}{n} \sum_{i=1}^n \nu_i \Big[ T_i (Y_i - \mu_{IPW,1})^2 + (1-T_i)(Y_i - \mu_{IPW,0})^2 \Big]
\end{align}


\subsection{NN and caliper matching}
When utilizing NN or caliper matching with BPSA, the conditional distribution of $\Delta$ is estimated asymptotically. 

Both nearest neighbor and caliper matching were performed with the \verb|MatchIt| package on \verb|R|. $\nu_i$, defined as the frequency weight outputted by the \verb|MatchIt| object (see Section \ref{ps.design:imp:nn}), is utilized in a weighted regression of the same model as Equation \ref{eq:msm}. Estimation of $E(\hat{\Delta}^*(\nu))$, $Var(\hat{\Delta}^*(\nu))$ (as utilized in BPSA) and $\hat{Var(\hat{\Delta})}$ (as utilized in PSA) proceed as described in Appendix \ref{apx:ipw} where frequency weights as described in the \verb|MatchIt| package documentation \cite{matchit} replace IP weights for $\nu$.

\newpage
\section{Tables and Figures}


\begin{center}
\begin{table}[b]%
\caption{Simulation result table for BPSA
\label{table:sim}}%
\scalebox{0.75}{%
\begin{tabular*}{300pt}{rllrrrrrrrr}
  \toprule
 Implementation & PS model & Empirical var & Avg total var & Avg between var & Avg within var & Bias & $PROP_{DU}$ & Coverage & MSE \\ 
  \midrule
Caliper matching (1-1) & confound & 0.015 & 0.021 & 0.005 & 0.016 & 0.070 & 0.241 & 0.966 & 0.020 \\ 
  Caliper matching (1-1) & instru & 0.035 & 0.027 & 0.008 & 0.020 & 0.091 & 0.282 & 0.890 & 0.043 \\ 
  Caliper matching (1-1) & instru\_prog & 0.023 & 0.035 & 0.015 & 0.019 & 0.092 & 0.429 & 0.964 & 0.032 \\ 
  Caliper matching (1-1) & instru\_prog\_noise & 0.023 & 0.034 & 0.015 & 0.019 & 0.103 & 0.420 & 0.948 & 0.034 \\ 
  Caliper matching (1-5) & confound & 0.015 & 0.017 & 0.001 & 0.011 & 0.070 & 0.075 & 0.926 & 0.020 \\ 
  Caliper matching (1-5) & instru & 0.035 & 0.023 & 0.004 & 0.013 & 0.091 & 0.155 & 0.862 & 0.043 \\ 
  Caliper matching (1-5) & instru\_prog & 0.024 & 0.031 & 0.011 & 0.013 & 0.092 & 0.353 & 0.946 & 0.032 \\ 
  Caliper matching (1-5) & instru\_prog\_noise & 0.023 & 0.030 & 0.010 & 0.013 & 0.103 & 0.342 & 0.938 & 0.033 \\ 
  DR & confound & 0.014 & 0.020 & 0.000 & 0.019 & 0.001 & 0.012 & 0.960 & 0.014 \\ 
  DR & instru & 0.066 & 0.091 & 0.009 & 0.076 & 0.010 & 0.082 & 0.966 & 0.066 \\ 
  DR & instru\_prog & 0.054 & 0.134 & 0.017 & 0.108 & -0.012 & 0.055 & 0.998 & 0.054 \\ 
  DR & instru\_prog\_noise & 0.053 & 0.143 & 0.016 & 0.115 & -0.003 & 0.072 & 1.000 & 0.053 \\ 
  IPW & confound & 0.016 & 0.015 & 0.004 & 0.011 & 0.002 & 0.252 & 0.938 & 0.016 \\ 
  IPW & instru & 0.068 & 0.023 & 0.012 & 0.011 & -0.001 & 0.438 & 0.790 & 0.068 \\ 
  IPW & instru\_prog & 0.079 & 0.036 & 0.026 & 0.011 & -0.036 & 0.646 & 0.892 & 0.080 \\ 
  IPW & instru\_prog\_noise & 0.084 & 0.039 & 0.028 & 0.011 & -0.045 & 0.668 & 0.886 & 0.086 \\ 
  NN matching & confound & 0.019 & 0.027 & 0.009 & 0.017 & 0.017 & 0.349 & 0.988 & 0.019 \\ 
  NN matching & instru & 0.062 & 0.047 & 0.024 & 0.023 & 0.021 & 0.473 & 0.910 & 0.063 \\ 
  NN matching & instru\_prog & 0.042 & 0.058 & 0.036 & 0.023 & 0.026 & 0.573 & 0.978 & 0.043 \\ 
  NN matching & instru\_prog\_noise & 0.041 & 0.056 & 0.033 & 0.023 & 0.037 & 0.567 & 0.962 & 0.042 \\ 
  Stratification & confound & 0.013 & 0.013 & 0.001 & 0.012 & 0.078 & 0.042 & 0.902 & 0.019 \\ 
  Stratification & instru & 0.029 & 0.034 & 0.005 & 0.029 & 0.100 & 0.133 & 0.926 & 0.039 \\ 
  Stratification & instru\_prog & 0.016 & 0.041 & 0.012 & 0.029 & 0.091 & 0.290 & 0.994 & 0.024 \\ 
  Stratification & instru\_prog\_noise & 0.016 & 0.042 & 0.012 & 0.029 & 0.106 & 0.285 & 0.984 & 0.027 \\ 
   \bottomrule
\end{tabular*}}
\end{table}
\end{center}

\begin{center}
\begin{table}[b]%
\caption{Simulation result table for PSA
\label{table:psa}}%
\scalebox{0.75}{%
\begin{tabular*}{300pt}{llrrrrrrrr}
  \toprule
 Implementation & PS model & Empirical var & Avg total var & Bias & Coverage & MSE \\ 
  \midrule
Caliper matching (1-1) & confound & 0.021 & 0.023 & 0.060 & 0.930 & 0.024 \\ 
Caliper matching (1-1) & instru & 0.042 & 0.043 & 0.084 & 0.930 & 0.049 \\ 
 Caliper matching (1-1) & instru\_prog & 0.032 & 0.041 & 0.079 & 0.942 & 0.038 \\ 
Caliper matching (1-1) & instru\_prog\_noise & 0.030 & 0.042 & 0.082 & 0.946 & 0.037 \\ 
Caliper matching (1-5) & confound & 0.017 & 0.019 & 0.061 & 0.926 & 0.021 \\ 
Caliper matching (1-5) & instru & 0.038 & 0.039 & 0.080 & 0.922 & 0.044 \\ 
Caliper matching (1-5) & instru\_prog & 0.027 & 0.038 & 0.077 & 0.940 & 0.033 \\ 
Caliper matching (1-5) & instru\_prog\_noise & 0.028 & 0.038 & 0.085 & 0.932 & 0.036 \\ 
DR & confound & 0.014 & 0.018 & 0.001 & 0.954 & 0.014 \\ 
DR & instru & 0.053 & 0.059 & 0.009 & 0.924 & 0.053 \\ 
DR & instru\_prog & 0.033 & 0.078 & -0.008 & 0.990 & 0.033 \\ 
 DR & instru\_prog\_noise & 0.033 & 0.078 & -0.001 & 0.984 & 0.033 \\ 
IPW & confound & 0.016 & 0.018 & 0.012 & 0.954 & 0.016 \\ 
IPW & instru & 0.060 & 0.052 & 0.028 & 0.916 & 0.061 \\ 
IPW & instru\_prog & 0.066 & 0.056 & 0.009 & 0.942 & 0.066 \\ 
IPW & instru\_prog\_noise & 0.069 & 0.060 & 0.008 & 0.938 & 0.069 \\ 
 NN matching & confound & 0.025 & 0.031 & 0.012 & 0.964 & 0.025 \\ 
NN matching & instru & 0.088 & 0.080 & 0.011 & 0.928 & 0.088 \\ 
NN matching & instru\_prog & 0.065 & 0.076 & 0.015 & 0.960 & 0.066 \\ 
NN matching & instru\_prog\_noise & 0.074 & 0.075 & 0.012 & 0.934 & 0.074 \\ 
 Stratification & confound & 0.013 & 0.013 & 0.067 & 0.906 & 0.017 \\ 
Stratification & instru & 0.034 & 0.031 & 0.084 & 0.912 & 0.041 \\ 
Stratification & instru\_prog & 0.021 & 0.031 & 0.071 & 0.980 & 0.026 \\ 
 Stratification & instru\_prog\_noise & 0.021 & 0.032 & 0.084 & 0.962 & 0.028 \\ 
   \bottomrule
\end{tabular*}}
\end{table}
\end{center}

\end{document}